\documentclass[aps,twocolumn,floatfix,showpacs,superscriptaddress,prb,eqsecnum]{revtex4}
\usepackage{graphicx}
\usepackage{amsmath}
\usepackage{amssymb}
\usepackage{bm}
\usepackage{color}

\begin{document}

\title{Metal--topological-insulator transition in the quantum kicked rotator with $\bm{\mathbb{Z}_{2}}$ symmetry}
\author{E. P. L. van Nieuwenburg}
\affiliation{Instituut-Lorentz, Universiteit Leiden, P.O. Box 9506, 2300 RA Leiden, The Netherlands}
\author{J. M. Edge}
\affiliation{Instituut-Lorentz, Universiteit Leiden, P.O. Box 9506, 2300 RA Leiden, The Netherlands}
\author{J. P. Dahlhaus}
\affiliation{Instituut-Lorentz, Universiteit Leiden, P.O. Box 9506, 2300 RA Leiden, The Netherlands}
\author{J. Tworzyd\l o}
\affiliation{Institute of Theoretical Physics, Faculty of Physics, University of Warsaw, \\ Ho\.{z}a 69, 00--681 Warsaw, Poland}
\author{C. W. J. Beenakker}
\affiliation{Instituut-Lorentz, Universiteit Leiden, P.O. Box 9506, 2300 RA Leiden, The Netherlands}
\date{January 2012}
\begin{abstract}
The quantum kicked rotator is a periodically driven dynamical system with a metal-insulator transition. We extend the model so that it includes phase transitions between a metal and a \textit{topological} insulator, in the universality class of the quantum spin Hall effect. We calculate the $\mathbb{Z}_{2}$ topological invariant using a scattering formulation that remains valid in the presence of disorder. The scaling laws at the phase transition can be studied efficiently by replacing one of the two spatial dimensions with a second incommensurate driving frequency. We find that the critical exponent does not depend on the topological invariant, in agreement with earlier independent results from the network model of the quantum spin Hall effect.
\end{abstract}
\maketitle

\section{Introduction}
\label{sec:intro}

The spin of an electron moving in an electric field experiences a torque, which can be understood as arising from the magnetic field in its rest frame. In a two-dimensional electron gas this velocity-dependent magnetic field produces the quantum spin Hall effect,\cite{Kan05,Koe07} reminiscent of the quantum Hall effect but without time-reversal symmetry breaking.\cite{Jac11} The difference manifests itself in the integer quantized values $Q$ of the dimensionless conductance. While there is no restriction on $Q\in\mathbb{Z}$ in the quantum Hall effect (QHE), only the two values $Q=0,1\in\mathbb{Z}_{2}$ appear in the quantum spin Hall effect (QSHE). In both effects the current is carried by edge states, separated by an insulating bulk. The insulator in the QSHE is called a topological insulator\cite{Has10,Qi11} if the topological quantum number $Q=1$ and a trivial insulator if $Q=0$.

Bulk states delocalize when the conductance switches between quantized values. A distinguishing feature of the QSHE is that the delocalized states can support metallic conduction (conductance $\gg e^{2}/h$), while in the QHE the conductance remains $\lesssim e^{2}/h$. The metallic conduction appears in extended regions of phase space, separated by a quantum phase transition (Anderson transition) from the regions with a quantized conductance and an insulating bulk. For a trivial insulator this is the familiar metal-insulator transition in a two-dimensional (2D) electron gas with spin-orbit scattering.\cite{Hik80,Eve08}

A fundamental question raised by the discovery of the QSHE was whether the phase transition would depend on the topological quantum number. Specifically, is the critical exponent $\nu_{Q}$ of the diverging localization length different if the phase boundary separates a metal from a topological insulator, rather than a trivial insulator? A numerical simulation\cite{Ono07} of the Kane-Mele model\cite{Kan05} of the QSHE gave an affirmative answer, finding a value $\nu_{1}\approx 1.6$ substantially below the established result\cite{Asa04,Mar06} for the metal--trivial-insulator transition. In contrast, studies of the network model in the QSHE universality class gave $\nu_{1}=\nu_{0}\approx 2.7$ within numerical accuracy,\cite{Obu07,Kob11} consistent with analytical considerations from the nonlinear sigma model\cite{Ryu10} why the critical exponent should be $Q$-independent.

In this work we study the metal-insulator transition in the QSHE by means of an altogether different, \textit{stroboscopic} model --- the periodically driven system known as the quantum kicked rotator.\cite{Cas79,Fis82,Sch89} A key feature of this dynamical system is that spatial dimensions can be exchanged for incommensurate driving frequencies,\cite{She83,Cas89} allowing for the study of metal-insulator transitions in one spatial dimension.\cite{Lem10,Tia11} This proved very effective in the QHE,\cite{Dah11} and also made it possible to experimentally study the 3D Anderson transition in a 1D optical lattice.\cite{Cha08} In the present paper we apply the same strategy to study the 2D QSHE in a 1D system.

In the next section we show how the quantum kicked rotator can be extended to include the topological $Q=1$ phase of the QSHE. We first construct this $\mathbb{Z}_{2}$ quantum kicked rotator in 2D and then carry out the mapping to 1D. We calculate the phase diagram in Sec.\ \ref{sec:TIdisorder}, using a scattering formula for the topological quantum number valid for disordered systems.\cite{Ful11,Mei11} In Sec.\ \ref{ssec:scaling} we determine the scaling law at the metal-insulator phase transition and compare the critical exponents $\nu_{Q}$ for $Q=0$ and $Q=1$. We conclude in Sec.\ \ref{sec:disc}.

\section{Construction of the $\bm{\mathbb{Z}_{2}}$ quantum kicked rotator}
\label{formulation_model}

\subsection{Stationary model without disorder}

We start from a translationally invariant 2D system and will add disorder later. The minimal model Hamiltonian $H_{0}(\bm{p})$ of the QSHE has four bands at each momentum $\bm{p}=(p_{1},p_{2})$, distinguished by indices $\sigma$ (up and down spins) and $\tau$ (\textit{s} and \textit{p} orbitals). The Pauli matrices $\sigma_{i}$ and $\tau_{i}$ ($i=0,x,y,z$) act on the spin and orbital degrees of freedom, respectively. Time-reversal symmetry is essential,
\begin{equation}
\sigma_{y}H^{\ast}(-\bm{p})\sigma_{y}=H(\bm{p}).\label{TRS}
\end{equation}
Inversion symmetry is not essential (and will be broken anyway once we add disorder), but is assumed for convenience,
\begin{equation}
\tau_{z}H(-\bm{p})\tau_{z}=H(\bm{p}).\label{inversion}
\end{equation}

The generic Hamiltonian that satisfies the symmetries \eqref{TRS} and \eqref{inversion} has the form\cite{Mur07}
\begin{align}
&H(\bm{p})=E_{0}(\bm{p})+\sum_{\alpha=1}^{5}f_{\alpha}(\bm{p})\Gamma_{\alpha},\label{HGamma}\\
&\bm{\Gamma}=\bigl(\tau_x\sigma_z,\;\tau_y\sigma_0, \; \tau_z\sigma_0, \;\tau_x\sigma_x, \; \tau_x\sigma_y \bigr).\label{Gammadef}
\end{align}
The real functions $E_{0},f_{3}$ are even under inversion of $\bm{p}$, while the functions $f_{1},f_{2},f_{4},f_{5}$ are odd. Because the $\Gamma$-matrices anticommute, $\{\Gamma_{\alpha},\Gamma_{\beta}\}=2\delta_{\alpha\beta}$, the band structure is given by
\begin{equation}
\varepsilon_{\pm}(\bm{p})=E_{0}(\bm{p})\pm\sqrt{{\textstyle\sum_{\alpha}}f_{\alpha}^{2}(\bm{p})}.\label{Epmdef}
\end{equation}
Each band is twofold degenerate.

The band gap can close upon variation of a single control parameter at high-symmetry points $\bm{\Lambda}_{a}$ in the first Brillouin zone, which satisfy $\bm{\Lambda}_{a}=-\bm{\Lambda}_{a}$ modulo a reciprocal lattice vector. At these time-reversally invariant momenta the Bloch wave function $u_{-}(\bm{\Lambda}_{a})$ of the lower band has a definite parity $\pi_{a}=\pm 1$ under inversion, $\tau_{z}u_{-}(\bm{\Lambda}_{a})=\pi_{a}u_{-}(\bm{\Lambda}_{a})$. The $\mathbb{Z}_{2}$ topological quantum number $Q$ follows from the Fu-Kane formula,\cite{Fu07}
\begin{equation}
(-1)^{Q}=\prod_{a}\pi_{a}.\label{Qpidef}
\end{equation}
A gap closing and reopening can switch the parity of the lower and upper bands, inducing a change in $Q$ (a topological phase transition).

The specific choice for the functions $f_{\alpha}(\bm{p})$ which we will study in the following is based on experience with the stroboscopic model of the QHE.\cite{Dah11} We take
\begin{equation}
E_{0}\equiv 0,\;\; f_{\alpha}=T(u)u_{\alpha},\;\;T(u)=(2/u)\,{\rm arctan}\,u,\label{E0Tu}
\end{equation}
where the vector $\bm{u}$ (of length $u=|\bm{u}|$) has components
\begin{align}
\bm{u}(\bm{p})={}&\biglb( K\sin p_1, \; K\sin p_2, \; \beta K(\mu - \cos p_1 - \cos p_2),\nonumber\\
&\;\;\gamma K\cos p_1\sin p_2, \; \gamma K\cos p_2\sin p_1 \bigrb).\label{udef}
\end{align}

For $\gamma=0$ this is the Bernevig-Hughes-Zhang model of the QSHE,\cite{Ber06} up to a function $T(u) > 0$ which flattens the bands without closing the band gap (hence without affecting the topological quantum number). Without the $\gamma$-term the spin degree of freedom $\sigma$ is conserved and the QSHE Hamiltonian is identical to two copies of the QHE Hamiltonian (with opposite magnetic fields, to restore time-reversal symmetry). When $\gamma\neq 0$ spin-orbit coupling mixes the spin-up and spin-down blocks of the Hamiltonian.

\begin{figure}[tb] 
\centerline{\includegraphics[width=0.9\linewidth]{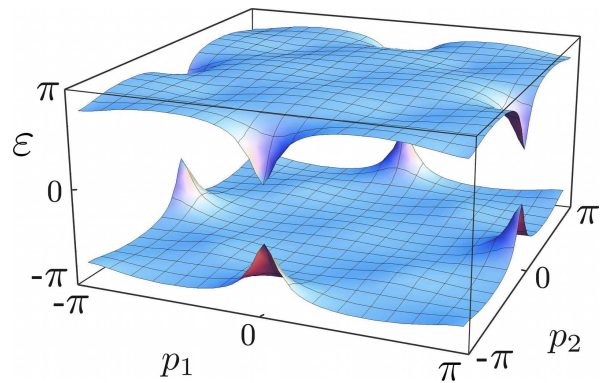}}
\caption{\label{fig:band} (Color online) 
Band structure for the clean Hamiltonian $H(\mathbf{p})$, as calculated from Eqs.\ \eqref{Epmdef}, \eqref{E0Tu}, and \eqref{udef}, for parameter values $K = 2$, $\beta = 0.8$, $\gamma = 2$, $\mu = -0.3$. The Dirac cones emerging at the time-reversally invariant points in the Brillouin zone will touch when $\mu = 0$.
}
\end{figure}

The band structure for one set of parameters is shown in Fig.\ \ref{fig:band}. The time-reversally invariant momenta $\bm{\Lambda_{a}}$ in the first Brillouin zone $-\pi<p_{1},p_{2}\leq\pi$ are at the four points $(0,0)$, $(\pi,\pi)$, $(0,\pi)$, $(\pi,0)$. (We have set both $\hbar$ and the lattice constant $a$ equal to unity.) The topological quantum number \eqref{Qpidef} depends only on the parameter $\mu$,
\begin{equation}
Q=\left\{\begin{array}{ll}
0&{\rm if}\;|\mu|>2,\\
1&{\rm if}\;|\mu|<2.
\end{array}\right.\label{Qclean}
\end{equation}
For $\mu=2$ or $-2$ the band gap closes at $(0,0)$ or $(\pi,\pi)$, with a switch in parity and a change of $Q$. For $\mu=0$ the gap closes at both points $(0,\pi)$ and $(\pi,0)$ --- at constant $Q$ since the two parity switches cancel.

\subsection{Time-dependent model with disorder}
\label{sm}

The time-dependent model is based on the quantum kicked rotator\cite{Cas79}, which is a dynamical system designed to study the localization by disorder with great nuerical efficiency.\cite{Fis82, Sch89, Tia11} The time dependent Hamiltonian $\mathcal{H}(t)$ contains a disorder potential $V(\bm x)$ and a stroboscopic kinetic energy $H(\bm{p})$,
\begin{equation}\label{eq:stroboscopic}
	\mathcal{H}(t) = V(\mathbf{x}) + H(\mathbf{p})\sum_{n=-\infty}^{\infty}\delta(t-n).
\end{equation}
(The stroboscopic period $\tau$ has been set equal to unity.) We take $H(\bm{p})$ of the form \eqref{HGamma} and will specify $V(\bm{x})$ later. The disorder strength is set by the relative importance of $V(\bm{x})$ and $H(\bm{p})$. The Floquet operator $\mathcal{F}$ describes the time evolution of the wavefunction over one period, 
\begin{equation}\label{eq:floquet}
\Psi(t+1) = {\cal F}\Psi(t),\;\;{\cal F} = e^{-iH(\bm{p})}e^{-iV(i\partial_{\bm{p}})}.
\end{equation}
Here $i\partial_{\mathbf{p}}$ is the operator $\mathbf{x}$ in the momentum representation. The eigenvalues $e^{-i\varepsilon}$ of the unitary operator ${\cal F}$ define the quasi-energies $\varepsilon \in (-\pi,\pi)$. 

We use the $2\pi$-periodicity of $H(\bm{p})$ to label the eigenstates $\Psi_{\bm{q}}(\bm{p})$ of ${\cal F}$ by a Bloch vector $\bm{q}$ in the Brillouin zone $ -\pi < q_1, q_2 \leq \pi$. By construction,
\begin{equation}\label{eq:floqeigst}
	\Psi_{\bm{q}}(\bm{p}) = e^{-i\bm{p}\cdot\bm{q}}\chi_{\bm{q}}(\bm{p}),
\end{equation}
with $\chi_{\bm{q}}(\bm{p})$ a $2\pi$-periodic eigenstate of 
\begin{equation}\label{eq:floqbloch}
	{\cal F}_{\bm{q}} = e^{-iH(\bm{p})}e^{-iV(i\partial_{\bm{p}} + \bm{q})}.
\end{equation}
This is the quantum kicked rotator with $\mathbb{Z}_{2}$ symmetry.

\subsection{Mapping from 2D to 1D}
\label{ssec:1DMap}

By adding an incommensurate driving frequency to the quantum kicked rotator in two spatial dimensions, it is possible to simulate the system in one single dimension.\cite{She83,Cas89} For that purpose we take a potential of the form
\begin{equation}\label{eq:linpot}
V(\bm{x}) = V_1(x_1) - \omega x_2,
\end{equation}
with $\omega/2\pi$ an irrational number $\in(0,1)$. We consider states which at $t=0$ are plane waves in the $x_{2}$-direction, having a well-defined initial momentum $p_{2}=\alpha$. In the Floquet operator the term linear in $x_{2}$ has the effect of shifting $p_2$ to $p_2 + \omega$, so that the 2D operator ${\cal F}_{\bm{q}}$ can be replaced at the $n$-th time step by the 1D operator
\begin{equation}\label{eq:1Dfloquet}
\mathcal{F}_{q}^{(n)} = e^{-iH(p_1,\;n\omega + \alpha)}e^{-iV_1(i\partial_{p_1} + q)}.
\end{equation}
There is considerable freedom in the choice of the potential $V_1$ in the remaining dimension. The simple quadratic form
\begin{equation}\label{eq:v1}
V_1(x_1) = \lambda(x_1 - x_0)^2
\end{equation}
provides sufficient randomness if $\{\omega,\lambda,2\pi\}$ form an incommensurate triplet.\cite{Bor97} We take $\lambda=1$, $\omega = 2\pi/\sqrt{5}$.

The numerical simulation is performed by subsequent multiplication of the state $\chi_{q}(x_1,t)$ with the series of Floquet operators ${\cal F}_{q}^{(n)}$, $n=0,1,2,\ldots$, in the plane wave basis $e^{-imp_1}$ (eigenstates of $x_1$). The integer $m$ is restricted to the values $1,2,\ldots M$, with $M$ an even integer that sets the system size. As initial condition we choose
\begin{equation}\label{eq:initialstate}
\chi_q(x_1, 0) = \delta_{x_1, x_0}\bm{\Sigma},
\end{equation}
spatially localized at $x_0=M/2$. The vector $\bm{\Sigma}$ is a normalized vector of rank 4 with random components denoting the spin and orbital degrees of freedom.

The multiplication with the Floquet operators, represented by $M\times M$ unitary matrices, can be done very efficiently by means of the Fast Fourier Transform algorithm. (See the analogous calculation in the QHE for a more detailed exposition.\cite{Dah11}) The calculation is repeated for different values of $\alpha$ and $q$, to simulate a disorder average. The disorder strength can be varied by varying $K$, with \textit{small} $K$ corresponding to \textit{strong} disorder.

\section{Phase diagram with disorder}
\label{sec:TIdisorder}

Before embarking on a calculation of the scaling law and critical exponents, we first locate the metal-insulator transitions and identify the topological phases. 

To find the metal-insulator transitions, we calculate the time dependent diffusion coefficient
\begin{equation}\label{eq:diff}
D(t) = \frac{\Delta^2(t)}{t}
\end{equation}
from the mean squared displacement 
\begin{equation}\label{eq:deltasquared}
\Delta^2(t) = \overline{\bigl\langle \bigl(x_{1}(t)-x_{0}\bigr)^{2}\bigr\rangle}.
\end{equation}
The brackets $\langle\cdots\rangle$ denote the expectation value in the state $\chi_{q}(x_{1},t)$ and the overline indicates the ensemble average over $\alpha$ and $q$. We typically average over $10^3$ samples of size $M=4\cdot10^3$.

\begin{figure}[tb] 
\centerline{\includegraphics[width=0.9\linewidth]{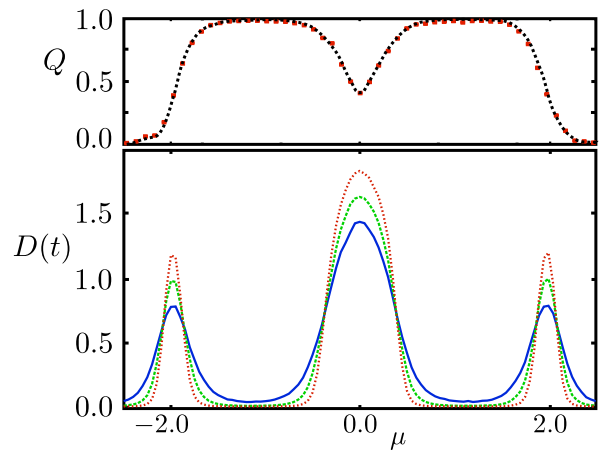}}
\caption{\label{fig:DandTscan} (Color online) 
\textit{Bottom panel:} Time-dependent diffusion coefficient \eqref{eq:deltasquared} as a function of $\mu$ for $K=2$, $\beta = 0.8$, $\gamma = 2$, shown for times $t=10^6$ (red), $t=10^5$ (green) and $t=10^4$ (blue). The points of intersection of these curves locate the metal-insulator transition. \textit{Top panel:} Topological quantum number \eqref{eq:TIdisordered} for the same parameter values, used to distinguish the topologically trivial ($Q=0$) and nontrivial ($Q=1$) insulators. In the metallic regions $Q$ is not quantized.
}
\end{figure}

\begin{figure}[tb] 
\centerline{\includegraphics[width=0.9\linewidth]{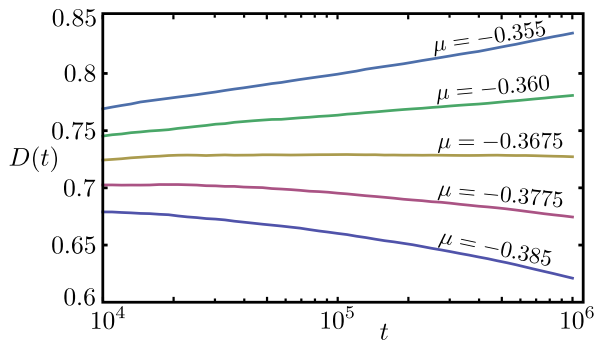}}
\caption{\label{fig:logscaling} (Color online) 
Time-dependent diffusion coefficient as a function of $t$ for several values of $\mu$ near the metal--topological-insulator transition. (Same parameters as in Fig.\ \ref{fig:DandTscan}.) The slope vanishes at the transition point $\mu_{c} \approx -0.368$.
}
\end{figure}

A representative series of scans of $D$ versus $\mu$ for different values of $t$ is shown in Fig.\ \ref{fig:DandTscan} (bottom panel). In Fig.\ \ref{fig:logscaling} we show $D$ versus $t$ for different $\mu$. In the insulating phase $D(t)\propto 1/t$ decays with increasing time, while in the metallic phase $D(t)\propto\ln t$ grows with increasing time.\cite{Asa06} The metal-insulator transition at $\mu=\mu_{c}$ is signaled by a $t$-independent $D(t)$ (corresponding to a scale invariant diffusion coefficient). 

In this way we can locate the phase boundaries, but we cannot yet distinguish topologically trivial and nontrivial insulators. For that purpose we need the topological quantum number. The formula \eqref{Qpidef} for the topological quantum number of the translationally invariant system does not apply for nonzero disorder potential. The scattering formulation\cite{Ful11,Mei11} continues to apply and is what we will use. An alternative Hamiltonian formulation for disordered systems has been given by Prodan.\cite{Pro11a,Pro11b}

The topological invariant is computed using the formalism described in Ref.\ \onlinecite{Dah11}.  From the Floquet operator we can construct a reflection matrix $r(\varepsilon,\phi)$  for a cylindrical system enclosing a flux  $\Phi=\phi\hbar/e$. The topological invariant then follows from the following combination of determinants and Pfaffians, evaluated at $\varepsilon=0$ and $\phi=0,\pi$,
\begin{equation}
(-1)^{Q}=\frac{\text{Pf}\,[\sigma_{y}r(0,\pi)]}{\text{Pf}\,[\sigma_{y} r(0,0)]} \frac{\sqrt{\det r(0,0)}}{\sqrt{\det r(0,\pi)}}.
\label{eq:TIdisordered}
\end{equation}

The results in Fig.\ \ref{fig:DandTscan} (top panel) show the disorder-averaged $\mu$-dependence of $Q$ for a system of size $M\times M=30\times 30$. The value of $Q$ is only quantized $\in\{0,1\}$ in the insulating regions. In the metallic regions $Q$ averages to $1/2$ for a sufficiently large system,\cite{Ful11,Pro11b} which is not quite observed for our system sizes.

\begin{figure}[tb] 
\centerline{\includegraphics[width=0.8\linewidth]{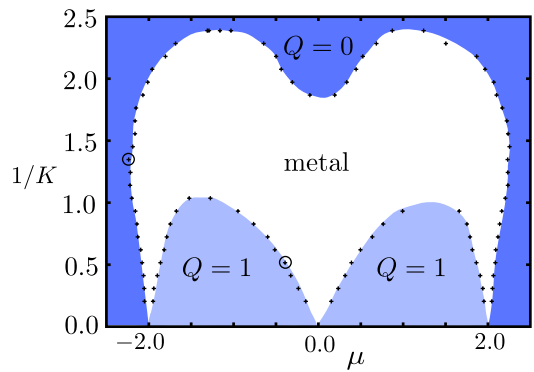}}
\caption{ \label{fig:pd} (Color online) 
Phase diagram of the $\mathbb{Z}_{2}$ quantum kicked rotator in the $\mu$--$1/K$ plane for fixed $\beta = 0.8$ and $\gamma = 2$. The topological quantum number $Q=0,1$ distinguishes the topologically trivial and nontrivial insulating phases. The metallic phase between the insulating phases disappears in the clean limit $1/K\rightarrow 0$. The two circles indicate the metal--trivial-insulator and metal--topological-insulator transitions that were studied to compare the critical exponents.
}
\end{figure}

\begin{figure}[tb] 
\centerline{\includegraphics[width=0.8\linewidth]{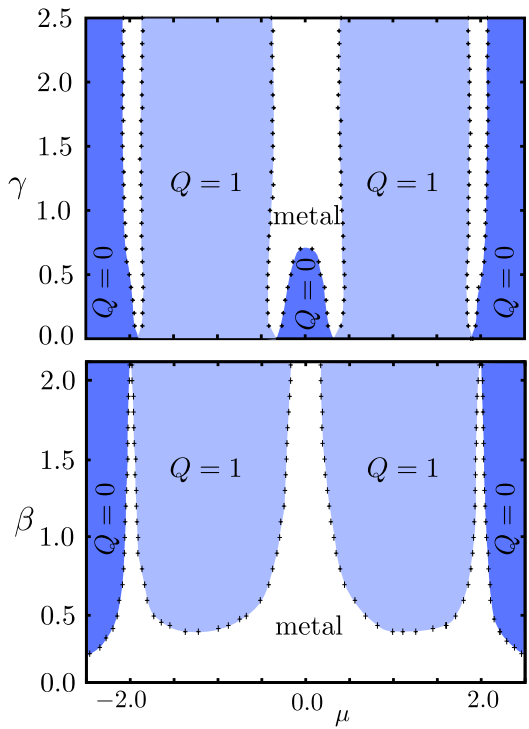}}
\caption{\label{fig:gammapd} (Color online) 
Phase diagram for fixed disorder strength ($K=2$), in the $\mu$--$\gamma$ plane for $\beta = 0.8$ (top panel) and in the $\mu$--$\beta$ plane for $\gamma = 2$ (bottom panel). The metallic phase is stabilized by increasing the spin-orbit coupling strength $\gamma$ or by decreasing $\beta$.
}
\end{figure}

The phase diagram obtained in this way is shown in Figs.\ \ref{fig:pd} and \ref{fig:gammapd}. Without disorder, the topological invariant \eqref{Qclean} gives a topological insulator for $|\mu|<2$ and a trivial insulator for $|\mu|>2$. With disorder a metallic phase appears between the insulating phases, provided that $\gamma/\beta\neq 0$.
For small $\gamma$ we find an additional trivial insulating phase around $\mu=0$, consistent with what was found in the quantum Hall system in the presence of disorder \cite{Dah11}.

In some other models of the QSHE a reentrant behavior $Q=0\mapsto 1\mapsto 0$ with increasing disorder is observed in some regions of parameter space.\cite{Pro11b,Li09,Gro09,Yam11} In our model a system which starts out topologically trivial in the clean limit stays trivial with disorder.

\section{Scaling law and critical exponent}
\label{ssec:scaling}

The premise of one-parameter scaling is expressed by the equation\cite{Mac83} 
\begin{equation}\label{eq:ops}
	\ln D(t) = {\cal F}(\xi^{-2}t),
\end{equation}
where ${\cal F}$ is a universal scaling function. The localization length $\xi$ has a power law divergence $\propto |\mu-\mu_{c}|^{-\nu}$ at the metal-insulator transition, with critical exponent $\nu$. 

We follow the established method of finite-size (here: finite-time) scaling to extract $\nu$ from the numerical data.\cite{Sle09} We rewrite the scaling law \eqref{eq:ops} as
\begin{align}
&\ln D(t) = F(ut^{1/2\nu}),\label{eq:ops2}\\
& u = (\mu - \mu_c) + \sum_{k=2}^{N_u} c_k \left( \mu - \mu_c \right)^k,\label{eq:ufit}
\end{align}
where $F(z)$ is an analytic function of $z=ut^{1/2\nu}$. By fitting the free parameters of the series expansion
\begin{equation}\label{eq:fitfunc}
	\ln D(t) = \ln D_c + \sum_{k=1}^{N_F} d_k \left(u t^{1/2\nu}\right)^k + S(t)
\end{equation}
to the data for $D$ as a function of $\mu$ and $t$, the critical exponent $\nu$ is obtained. The extra term $S(t)$ accounts for finite-time corrections to single-parameter scaling, of the form
\begin{equation}\label{eq:fitfunccorr}
	S(t) = t^{-y} \sum_{k=0}^{N_s} g_k \left( (\mu-\mu_c)t^{1/2\nu} \right)^k.
\end{equation}

We have considered times up to $t=10^6$ for system sizes up to $M=10^4$. The number of terms $N_u$, $N_F$, and $N_s$ in the series expansions \eqref{eq:ufit}--\eqref{eq:fitfunccorr} are systematically increased until the chi-square-value per degree of freedom ($\chi^2/\text{ndf}$) is approximately unity, see Table \ref{tab:critexps}. The calculation is carried out at the two points indicated by circles in Fig.\ \ref{fig:pd}, one a metal--trivial-insulator transition (giving $\nu_{0} = 2.67 \pm 0.09$) and the other a metal--topological-insulator transition (giving $\nu_{1} = 2.69 \pm 0.06$).

\begin{table}[tb]
\begin{center}
\begin{tabular}{l||ccccc}
 & $N_u$ & $N_F$ & $N_s$ & $\nu$ & $\chi^2/\text{ndf}$ \\
\hline\hline
$\nu_{0}$ & 2 & 3 & 2 & $2.67 \pm 0.09$ & $0.97$ \\
\hline
$\nu_{1}$ & 2 & 1 & 2 & $2.69 \pm 0.06$ & $0.92$ \\
\hline
\end{tabular}
\caption{\label{tab:critexps}
Results of the finite-time scaling analysis described in the text.
}
\end{center}
\end{table}

\section{Conclusion}
\label{sec:disc}

In conclusion, we have presented a numerical method to study the metal-insulator transition in the quantum spin Hall effect (QSHE), based on the incorporation of $\mathbb{Z}_{2}$ topological symmetry into the quantum kicked rotator. We find that the critical exponent $\nu_{Q}$ of the diverging localization length is the same whether the metal is approached from the topologically trivial insulator ($Q=0$) or from the topologically nontrivial insulator ($Q=1$). Our results $\nu_{0}=\nu_{1}\approx 2.7$ are in agreement with Refs.\ \onlinecite{Obu07,Kob11}, but not with Ref.\ \onlinecite{Ono07} (which found a much smaller $\nu_{1}\approx 1.6$ for the topologically nontrivial insulator). Since our stroboscopic model is fully independent of the network model used in Refs.\ \onlinecite{Obu07,Kob11}, this is significant support for the insensitivity of $\nu$ to the topological quantum number $Q$.

A special feature of the quantum kicked rotator is that it allows the study of the QSHE, as well as the QHE,\cite{Dah11} in one spatial dimension, by exchanging a dimension for an incommensurate driving frequency.\cite{She83,Cas89} There is much interest in such one-dimensional models of topological phases,\cite{Kit10,Lin11,Kra11,Mei11b} because they might be more easily realized in optical lattices of cold atoms than the original two-dimensional models. Critical exponents were not studied in these earlier investigations, which focused on the QSHE in clean systems without disorder.

An interesting direction for future research is to study the edge states in the $\mathbb{Z}_{2}$ quantum kicked rotator, by replacing the periodic boundary condition used in this work by a zero-current boundary condition. While the localization length exponent $\nu$ does not depend on the topological quantum number, the edge state structure does depend on $Q$, with a characteristic multifractality at the metal-insulator transition.\cite{Obu07}

\acknowledgments

This research was supported by the Dutch Science Foundation NWO/FOM, by an ERC Advanced Investigator Grant, and by the EU network NanoCTM.

\end{document}